\documentclass[aps,prl,twocolumn,showpacs,superscriptaddress,groupedaddress]{revtex4}

\usepackage{graphicx}  
\usepackage{dcolumn}   
\usepackage{bm}        
\usepackage{amssymb}   

\begin{document}
\widetext
\leftline{Version 1st as of \today}
\leftline{Hong-Wei TAN, Jin-Bo YANG, Tang-Mei HE, Jing-Yi ZHANG}
\leftline{To be submitted to (CPT)}
\leftline{Corresponding author: Jing-Yi ZHANG (\tt{ physicz@aliyun.com})}
\leftline{Center for Astrophysics, Guangzhou University, Guangzhou 510006}
\centerline{\em D\O\ INTERNAL DOCUMENT -- NOT FOR PUBLIC DISTRIBUTION}


\title{A modified thermodynamics method to generate exact
solutions of the Einstein equations\footnote{This research is supported by the National Natural Science Foundation of China under Grant Nos.11273009 and 11303006.}}
\date{\today}

\begin{abstract}
We modify the method to generate the exact solutions of the Einstein equations basing on the laws of thermodynamics. Firstly, the Komar mass is used  to take the place of the Misner-Sharp energy which is used in the original methods, and then several exact solutions of Einstein equations are obtained, including the black hole solution which surrounded by quintessence. Moreover, the geometry surface gravity defined by Komar mass is obtained. Secondly, we use both the Komar
mass and the ADM mass to modify such method, and the similar results are obtained. Moreover,
with some generalize added to the definition of the ADM mass, our method can be generalized to
global monopole sapcetime.
\end{abstract}

\maketitle


\section{Introduction}
Since Bekenstein found the relationship between black hole dynamics and thermodynamics \cite{beckenstein},
and Hawking presented Hawking radiation according to the quantum field theory in curved space-time \cite{hawking}, which is a pure thermodynamical radiation, people have focused on the deep relationship between the theory of the gravitation and thermodynamics for a long time.

In fact, black hole thermodynamics can be viewed as spacetime thermodynamics, which
means that the properties of the physical objects in black hole thermodynamics is global on a manifold which is equipped with a Lorentz metric, known as a spacetime. However, it is
very difficult to construct thermodynamics in general situations for some common physical quantities such as mass, entropy and angular momentum which can not be well defined. Moveover, in a general space-time, the thermodynamics is usually need to be considered as an nonequilibrium state, which is very difficult to be dealt with even for
ordinary matter. Though there are such difficulties, it does not  stop the researchers from deriving the Einstein equations from thermodynamic laws, the problems in inverse logic \cite{zhang}.

In 1995, Jacobson derived the Einstein equations from the basic equations of thermodynamics and the Raychaudhuri equations on the null hypersurface \cite{jacobson}, by using the local first law of equilibrium thermodynamics. In such work, the researchers used the assumption that the entropy is proportional to the area of the local Rindler horizon of an infinitely accelerated observer, and the Hawking-Unruh temperature, which had been exploited in Ref. \cite{Unruh}, was treated as the
 temperature observed by such observer. Basing on such assumptions, the Einstein equations were derived. However, in that work, the researchers assumed that the spacetime is in a locally
thermal equilibrium system, but as the equations that describe the evolution of all spacetimes, Einstein equations are expected to be able to describe all kinds of  spacetimes' evolution in principle naturally, including the spacetimes that do not satisfy the locally thermal equilibrium assumption. In other words,  the researchers obtained the equations that can describe general situations only based on a special assumption, which is unnatural in logic, see \cite{zhang}.

For this reason, Ref.\cite{zhang} put forward a new method to study such problem. In such paper, the researchers  Considered a spacetime equipped with spherically symmetry, whose metric ansatz is $ds^2=-f(r)dt^2+h(r)dr^2+r^2d\Omega^2$. In such spacetime, the energy of the gravitational field was defined as the Misner--Sharp energy \cite{misner}. Firstly, the researchers applied the first law of equilibrium thermodynamics in a  adiabatic system, $dM=dW$, to derive $h(r)$. Deriving $f(r)$ is a difficult task, to solve such problem, the researchers assumed that the surface gravity defined in the traditional way is equal to the geometry surface defined by the unified first law \cite{hayward}, and then they generated several exact solutions of the Einstein equations. There is no doubt that the amazing results obtained in Ref.\cite{zhang} provides a new way to study the gravitational thermodynamics. However, there is a limitation in this method for such method requires the symmetry of the space-time strictly because the Misner--Sharp energy can only be defined in the spacetime with a spherically symmetry, a plane symmetry as well as a Pseudo spherically symmetry\cite{a,b,c}. This difficulty motivates us to modify this method.

  There are two steps of such modification introduced in our paper. Firstly, we replace the Misner--Sharp mass with only the Komar mass \cite{komar}, by using the first law of equilibrium thermodynamics in a  adiabatic system just like the original method did, and then the results obtained here are similar with that obtained in the original method. Since the definition of the Komar mass only requires that the space-time is stationary, means that if there is time-like  a Killing vector in the space-time, then our method can be used in principle. In addition, the
black hole solution surrounded by quintessence is also generated in this paper. Our another achievement is that we construct another definition of the geometry surface gravity, which is defined by the Komar mass. In the second step of our work, we use the ADM mass , another quasi-local energy for gravitational field \cite{ADM}, together with the Komar mass to complete such modification. If we do so, then we can also regenerate these exact solutions of Einstein equations, further more, we modify the definition of the ADM mass, and then the global monopole spacetime can be generated.

This paper is organized as follows: in section 2 we modify this method with only the Komar mass, and general severate exact solution of Einstein equations, the geometry surface gravity defined by Komar mass is also construct in this section; in section 3, we introduce the method modified by both the Komar mass and the ADM mass, and some comments on the situation that the spacetime with global monopole charge is arisen; in the  section 4 and 5, some discussion and conclusion is given.
\section{Modified with only Komar mass}
In this section, the method modified with only the Komar mass will be introduced. Here, the metric ansatz of a spherically symmetric spacetime is
\begin{equation}\label{sph}
ds^2=-f(r)dt^2+h^{-1}(r)dr^2+r^2d\Omega^2.
\end{equation}
In a stationary space-time, the Komar mass can be defined as
\begin{equation}\label{k2}
M_k=-\frac{1}{8\pi}\int_S\epsilon_{abcd}\nabla^c\xi^d,
\end{equation}
where $\epsilon_{abcd}$ is the volume element of the four dimensional space-time and $\xi^d$ is a time-like Killing vector field. According to the two formulas above, we can get the Komar mass in this metric ansatz as
\begin{equation}\label{k}
M_k=\frac{r^2}{2}\sqrt{\frac{h}{f}}\frac{df}{dr}.
\end{equation}

In the spherically symmetric space time, according to the unified first law,  we can define the geometry surface gravity as \cite{hayward}
\begin{equation}\label{kappag}
\kappa_g=\frac{M_{ms}}{r^2}-4\pi r\omega,
\end{equation}
in which $M_{ms}$ is the Misner-Sharp energy defined as \cite{misner}
\begin{equation}\label{ms}
M_{ms}=\frac{1}{2}r(1-h),
\end{equation}
 and $\omega$ is the work term defined as \cite{hayward}
\begin{equation}
\omega=-\frac{1}{2}I^{ab}T_{ab},
\end{equation}
where $I^{ab}$ is the inverse of the induced metric of the space-time in the leader two dimensions whose line element reads
\begin{equation}
I=-f(r)dt^2+h(r)dr^2.
\end{equation}
On the other hand, in Eq.(\ref{sph}), the surface gravity is
\begin{equation}\label{kappa}
\kappa=\frac{1}{2}\sqrt{\frac{h}{f}}\frac{df}{dr}.
\end{equation}
In Ref.\cite{zhang}, the researchers assumed that the surface gravity is equal to the geometry surface gravity
\begin{equation}\label{lappaeq}
\kappa=\kappa_g.
\end{equation}
In this paper we will follow this assumption. According to Eqs.(\ref{k}), (\ref{kappag}), (\ref{kappa}) and (\ref{lappaeq}), we can obtain the relationship between the Komar mass and the Misner-Sharp energy as
\begin{equation}\label{kme}
M_k=M_{ms}-4\pi\omega r^3.
\end{equation}
\subsection{The Schwarzschild solution}
Considering a vacuum spacetime and the first law of equilibrium thermodynamics in the adiabatic system, we can get
\begin{equation}\label{first}
dM_k=0.
\end{equation}
The energy-stress tensor is zero in the vacuum, so the work term $\omega$ must be zero. Combining Eqs.(\ref{ms}), (\ref{kme}) and (\ref{first}) together,
we have
\begin{equation}
d[\frac{1}{2}r(1-h)]=0.
\end{equation}
Solving this equation, the result reads
\begin{equation}
h=1-\frac{C}{r}.
\end{equation}
Substituting it into Eqs.(\ref{k}) and (\ref{ms}), and combing with eq.(\ref{kme}), $f(r)$ is obtained as
\begin{equation}\label{f}
f=[(1-\frac{C}{r})^{\frac{1}{2}}+D]^2.
\end{equation}
If we choose the asymptotically flat spacetime as our boundary condition, then
\begin{equation}
D=0.
\end{equation}
And the Komar mass reads
\begin{equation}\label{f2}
M_{k}=\frac{C}{2}.
\end{equation}
Finally the result can be written as
\begin{equation}
ds^2=-ги(1-\frac{2M}{r})dt^2+\frac{1}{1-\frac{2M}{r}}dr^2+r^2d\Omega^2.
\end{equation}
It is exactly the line element of the Schwarzschild space-time. Now we can draw a conclusion that the Kormar mass describes a adiabatic process. Furthermore, combining Eqs.(\ref{k}), (\ref{kappa}), (\ref{lappaeq}) together, we can obtain the surface gravity defined by the Komar mass as
\begin{equation}\label{nk}
\kappa^{\prime}=\frac{M_k}{r^2}.
\end{equation}
\subsection{The Schwarzschild-de Sitter solution}
Now let us deal with the situation that there is force works. Considering the first law of thermodynamics again
\begin{equation}\label{de}
dM_k=-PdV,
\end{equation}
where $P$ donates the pressure and $V$ is the volume
\begin{equation}\label{V}
V=\frac{4}{3}\pi r^3,
\end{equation}
the work term is \cite{zhang}
\begin{equation}\label{omega}
\omega=\frac{\Lambda}{8\pi}.
\end{equation}
Where $\Lambda$ can be viewed as the cosmological constant. Substituting it into Eqs.(\ref{kme}), then the Komar mass reads
\begin{equation}\label{k1}
M_k=M_{ms}-\frac{r^3\Lambda}{2}.
\end{equation}
Based on Eqs.(\ref{de}), (\ref{omega}), (\ref{k1}), we can get
\begin{equation}
d(\frac{r}{2}(1-h)-\frac{r^3\Lambda}{2})=-4\pi Pr^2dr.
\end{equation}
Letting the $\Lambda=4\pi P$, the results are read as
\begin{equation}\label{h2}
h=1-\frac{C}{r}-\frac{\Lambda r^2}{3},
\end{equation}
and
\begin{equation}
f=1-\frac{C}{r}-\frac{\Lambda r^2}{3}.
\end{equation}
It is just the line element of the  Schwarzschild de Sitter spacetime
\begin{equation}
ds^2=-\big(1-\frac{C}{r}-\frac{\Lambda}{3}r^2\big)dt^2+\frac{1}{1-\frac{C}{r}-\frac{\Lambda}{3}r^2}dr^2+r^2d\Omega^2.
\end{equation}
\subsection{The RN-de Sitter solution}
Furthermore, if we consider there is a electric charge, then
\begin{equation}
dM_k=-PdV+\frac{q^2}{r^2}dr.
\end{equation}
Ref.\cite{zhang} assumed that the work of the electric  field can be written as $\frac{q}{r}dq$, however, we find that this assumption can not derive the RN solution. Indeed, in Ref.\cite{hayward} the work of the electric force is considered as $\frac{q^2}{r^2}dr$ . Moreover, if we use this as our assumption and then the RN solution can be obtained, which will be expressed as follows.

To be more general, we consider that there are both force and electric field work, so the work term is written as \cite{zhang}
\begin{equation}
\omega=\frac{q^2}{8 \pi r^4}+\frac{\Lambda}{8\pi}.
\end{equation}
We can obtain the Komar mass in this situation as
\begin{equation}
  M_k=M_{ms}-\frac{r^3\Lambda}{2}-\frac{q^2}{2r}.
\end{equation}
So, we can get the equation as
\begin{equation}
d(\frac{r}{2}(1-h)-\frac{r^3\Lambda}{2}-\frac{q^2}{2r})=-4\pi Pr^2dr+\frac{q^2}{r^2}dr.
\end{equation}
Considering $\Lambda=4\pi P$ and solving the equation above, the solution is obtained as
\begin{equation}
h=1-\frac{C}{r}-\frac{\Lambda r^2}{3}+\frac{q^2}{r^2}.
\end{equation}
Substituting this into Eq.(\ref{k}), we get
\begin{equation}
f=1-\frac{C}{r}-\frac{\Lambda r^2}{3}+\frac{q^2}{r^2}.
\end{equation}
And therefore, we get the line element of RN-de Sitter spacetime
\begin{eqnarray}
ds^2&=&-\big(1-\frac{C}{r}+\frac{q^2}{r^2}-\frac{\Lambda}{3}r^2\big)dt^2\nonumber\\
    &+&\frac{1}{1-\frac{C}{r}+\frac{q^2}{r^2}-\frac{\Lambda}{3}r^2}dr^2+r^2d\Omega^2.
\end{eqnarray}
\subsection{More general situations}
In more general situations, if it is assumed that the work term $\omega$ and and the power $P$ are both power functions of $r$, apply the first law of thermodynamics, for convenient, the equation is expressed as
\begin{equation}
d(r(1-h)-cr^d)=ar^bdr,
\end{equation}
where $a, b, c$ and $d$ are all constants. The solution of this equation is
\begin{equation}
h=1-\frac{ar^b}{r^{b+1}}-cr^{d-1}-\frac{C_1}{r}
\end{equation}
If it is assumed that $a, b, c$ and $d$ are not independent with each other but constrained by following conditions
\begin{equation}\label{con}
d=1+b,  c=\frac{a}{1+d},
\end{equation}
then $h$ can be rewritten as
\begin{equation}
h=1-(\frac{a}{d}-\frac{a}{1+d})r^{d-1}-\frac{C_1}{r},
\end{equation}
Above formula can be inserted into Eq.(\ref{kme}), then we have
\begin{eqnarray}
&&r\{1-[(\frac{a}{d}-\frac{a}{1+d})r^{d-1}-\frac{C_1}{r}]\}-\frac{a}{1+d}r^d\nonumber\\
&=&r^2\sqrt{\frac{1-(\frac{a}{d}-\frac{a}{1+d})r^{d-1}-\frac{C_1}{r}}{f}}\frac{df}{dr},
\end{eqnarray}
and the solution is
\begin{eqnarray}
f&=&1-(\frac{a}{d}-\frac{a}{1+d})r^{d-1}-\frac{C_1}{r}\nonumber\\
&+&C_2[\frac{\sqrt{-ar^d+(d+d^2)(r+C_1)}}{\sqrt{rd(d+1)}}+\frac{C_2}{4}].
\end{eqnarray}
Setting $C_2=0$, so
\begin{equation}
f=1-(\frac{a}{d}-\frac{a}{1+d})r^{d-1}-\frac{C_1}{r}.
\end{equation}
Redefining a new parameter $\alpha$ as
\begin{equation}
\alpha=-(\frac{a}{d}-\frac{a}{1+d}),
\end{equation}
then the line element of the spacetime is
\begin{eqnarray}
ds^2&=&-\big(1-\frac{C_1}{r}+\frac{\alpha}{r^{1-d}}\big)dt^2\nonumber\\
    &+&\frac{1}{1-\frac{C_1}{r}+\frac{\alpha}{r^{1-d}}}dr^2+r^2d\Omega^2.
\end{eqnarray}
Where $\alpha$ can be viewed as the charge of the spacetime. For some specific examples, if $\alpha=0$, then $a=c=0$, the Schwarzschild solution can be obtained, and if $d=-1$, then we arrive at the RN spacetime and $\alpha=q^2$, where $q$ is the electric charge.

Noted that if we set the range of $d$ as
\begin{equation}
1<d<3,
\end{equation}
then we arrive at the black hole solution surrounded by quintessence, which has been obtained by V.V.Kiselev in 2003 \cite{Kiselev}.

It should be careful that when $d=1$. In such situation the solution is
\begin{equation}
f=h=1-\frac{C_1}{r}+\alpha,
\end{equation}
it seems that the global monopole is generated. However, if we submit above into Eq.(\ref{k}), then the Kormas can be obtained as
\begin{equation}
M=\frac{C_1}{2},
\end{equation}
and then the thermodynamical relationship reads
\begin{equation}
dM_k=0,
\end{equation}
which means that there is not any work in this situation. It requires that
\begin{equation}
a=c=0,
\end{equation}
in Eq.(\ref{con}), then we just arrive at the Schwarzschild situation again.
\section{MODIFIED WITH both KOMAR Mass and ADM mass}
In a asymptotically flat sapcetime, the ADM mass can be defined as \cite{ADM}
\begin{equation}\label{ADM}
M_{ADM}=\frac{1}{16\pi}\lim_{r\to\infty}\int_S(\partial_jh_{ij}-\partial_ih_{jj})N^idS.
\end{equation}
where the $h_{ij}$ is the spatial component of the induced metric in the asymptotically Descartes coordinates. In our spacetime metric ansatz, the line element of the induced metric can be written as
\begin{eqnarray}
d\hat{s}&=&h^{-1}(r)dr^2+r^2d\Omega^2\nonumber\\
&=&h(r)^{-1}[dr^2+h(r)r^{2}(d\theta^2+\sin^2\theta d\phi^2)].
\end{eqnarray}
    Since what we consider now is a asymptotically flat sapcetime, so it can be believed that
\begin{equation}
\lim_{r\to\infty}h(r)=1.
\end{equation}
So the spatial line element can be written approximately as
\begin{equation}\label{reduce}
d\hat{s}\approx h(r)^{-1}[dr^2+r^{2}(d\theta^2+\sin^2\theta d\phi^2)].
\end{equation}
After some calculations, the ADM mass can be written as
\begin{equation}\label{ADM1}
M_{ADM}=\lim_{r\to\infty}\frac{1}{2}h(r)^{-\frac{3}{2}}r^2\frac{dh(r)}{dr}.
\end{equation}
After the limitation has been taken, then
\begin{equation}\label{ADM1}
M_{ADM}=\lim_{r\to\infty}\frac{1}{2}r^2\frac{dh(r)}{dr}.
\end{equation}
\subsection{The Schwarzschild solution}
Applying again the first law of thermodynamics in a vacuum spacetime which is in an adiabatic system
\begin{equation}\label{1}
dM_{ADM}=0,
\end{equation}
Then the following differential equation can be obtained
\begin{equation}
r^2\frac{d^2h(r)}{dr^2}+2r\frac{dh(r)}{dr}=0.
\end{equation}
Solving this equation, the result reads
\begin{equation}
h=C_1-\frac{C_2}{r}.
\end{equation}
The condition of asymptotic flat spacetime requires that $C_1=1$. submitting it into Eq.(\ref{ADM1}),then  the result can be obtained as
\begin{equation}\label{h}
h=1-\frac{2M_{ADM}}{r}.
\end{equation}
Inserting above result into the Komar mass (\ref{k}) and using the first law of thermodynamic
\begin{equation}
dM_k=0,
\end{equation}
then the following equation can be obtained
\begin{equation}
d\large(\sqrt{\frac{1-\frac{2M_{ADM}}{r}}{f}}\frac{df}{dr}r^2\large)=0.
\end{equation}
Solving the above equation, the result is
\begin{eqnarray}
f(r)&=&e^{-2\arctan(\frac{2M_{ADM}\sqrt{r}C_1'}{\sqrt{2M_{ADM}-r}})}\nonumber\\
&\times &(1-\frac{2M_{ADM}}{r}+4M_{ADM}^2(C_1')^2)C_2'.
\end{eqnarray}
If the integral constants are chosen as
\begin{equation}
C_1'=0, C_2'=1.
\end{equation}
Then the solution can be written as follows
\begin{equation}\label{f}
f(r)=1-\frac{2M_{ADM}}{r}.
\end{equation}
Combining Eqs, (\ref{k}), (\ref{h}) and (\ref{f}) together,  then the following result is obtained
\begin{equation}
M_{ADM}=M_{k}.
\end{equation}
This result suggests that our method is reasonable. Applying this result into eq.(\ref{sph}), then the Schwarzschild solution can be obtained
\begin{equation}
ds^2=-(1-\frac{2M}{r})dt^2+\frac{1}{1-\frac{2M}{r}}dr^2+r^2d\Omega^2
\end{equation}
\subsection{A comment on the spacetime with a global monopole charge}
Let us consider eq.(\ref{ADM1}) again, which is under the condition that
\begin{equation}
\lim_{r\to\infty}h(r)=1,
\end{equation}
now let us assume that Eq.(\ref{ADM1}) still works in the spacetime with a global monopole charge. However, such spacetime is not a spherically symmetry spacetime anymore. Specifically, considering a global monopole spacetime, whose line element is
\begin{equation}
ds^2=-(1-\eta^2-\frac{2M}{r})dt^2+\frac{1}{1-\eta^2-\frac{2M}{r}}dr^2+r^2d\Omega^2.
\end{equation}
Where $\eta$ is a constant. This line element can be rescaled as
\begin{equation}
ds^2=-(1-\frac{2M}{r})dt^2+\frac{1}{1-\frac{2M}{r}}dr^2+(1-8\pi\eta^2)r^2d\Omega^2.
\end{equation}
In this spacetime, the integral $\int dS$ is not $4\pi$ but $4\pi(1-8\pi\eta^2)$, see Ref.\cite{global}. So, in order to carry the information of the global charge, we  define the ADM mass in such spacetime  as
\begin{equation}\label{ADM2}
M_{ADM}=\lim_{r\to\infty}\frac{1-8\pi\eta^2}{2}r^2\frac{dh(r)}{dr}.
\end{equation}
It should be noted that when $\eta=0$, the definition above reduces to Eq.(\ref{ADM1}).  Now we have generalized the definition of ADM mass in the spacetime with a global monopole charge, and let us call this mass as quasi ADM mass.

Now we are ready to explore  what will such generalization will give us. Firstly, we consider a global monopole spacetime.   By using the definition of the quasi ADM mass, then
\begin{equation}
dM=0.
\end{equation}
Next let us consider the global monopole spacetime with a electric  charge, whose line element is
\begin{eqnarray}
ds^2&=&-(1-\eta^2-\frac{2M}{r}+\frac{q^2}{r^2})dt^2\nonumber\\
&+&\frac{1}{1-\eta^2-\frac{2M}{r}+\frac{q^2}{r^2}}dr^2+r^2d\Omega^2.
\end{eqnarray}
With calculation, the thermodynamical relationship can be obtain as
\begin{equation}
dM_{ADM}=(1-8\pi\eta^2)\frac{q^2}{r^2}dr.
\end{equation}
This result means that the global monopole charge results in a correction factor in the thermodynamical relationship.

 Whatever, It is obvious that the first law of thermodynamics can be obtained in our definition of quasi ADM mass, which suggests that such generalization is reasonable.

\section{discussion }
There are several comments on our work introduced as follows:

1.  In section 3, we introduced the method that modified by both the Komar mass and the ADM mass. However, to be honest, only the Schwarzschild solution has been generated completely in our work. However, with some trick, some other exact solutions can also be regenerated. Let us take the RN solution as an example. Firstly, let us consider the thermodynamical relationship for ADM mass in this situation
\begin{equation}
dM_{ADM}=\frac{q^2}{r^2}dr,
\end{equation}
and the solution reads
\begin{equation}\label{h1}
h(r)=1-\frac{2M}{r}+\frac{q^2}{r^2}.
\end{equation}
Submitting above into Eq.(\ref{k}), and using the same thermodynamical relationship, then we have
\begin{equation}
d\large(\sqrt{1-\frac{2M}{r}+\frac{q^2}{r^2}}\frac{df}{dr}r^2\large)=\frac{q^2}{r^2}dr,
\end{equation}
the above equation is too difficult to be solved, but we can check that the following is one particular solution of this equation:
\begin{equation}
f(r)=1-\frac{2M}{r}+\frac{q^2}{r^2}.
\end{equation}
Here, the RN spacetime is generated though this trick is not strict enough.

2. Some analyses about the situation that the spacetime with global monopole charge are also given in section 3. However, we can go to the inverse logic. We assume that the thermodynamical relationship also works in this situation. In vacuum, the thermodynamical relationship reads
\begin{equation}
dM_{ADM}=0.
\end{equation}
and the solution reads
\begin{equation}
h(r)=C_1-\frac{2M}{r}.
\end{equation}
in this situation, the requirement of the asymptotically flat sapcetime is loosen, so the integral constant can be chosen as $C_1=1-\eta$, and the result reads
\begin{equation}
h(r)=1-\eta^2-\frac{2M}{r}.
\end{equation}
and the $f(r)$ can also be solved as
\begin{equation}
f(r)=1-\eta^2-\frac{2M}{r}.
\end{equation}
Then the global monopole spacetime has been generated.
\section{Conclusion}
In this paper, we modify the method to generate the exact solution of the Einstein equations with the laws of thermodynamics which was arisen in Ref.\cite{zhang}. In Ref.\cite{zhang}, the researchers used the Misner-Sharp energy and unified first law to derive several exact solutions of Einstein equations without involving it. However, the Misner-Sharp energy can only be defined
in the space-time with a spherically symmetry, a plane symmetry as well as a Pseudo spherically symmetry, which limits this method to be generalized to more general situation.

This method is modified in two steps in this paper. Firstly, we use only the Komar mass to take the place of the Misner-Sharp energy to modify such method, and then  several exact solutions of the Einstein equations are regenerated. Moreover, we obtain the geometry surface gravity defined by the Komar mass in the specially symmetry space-time.
Since the Komar mass requires the symmetry less than the Misner-Sharp energy, which results in our method could be used in more situations general in principle.

Secondly, we modify this method with both the Komar mass and the ADM mass, some exact solutions of  Einstein can also be regenerated. Moreover, the quasi ADM mass is defined in the spacetime with a global monopole charge   and some thermodynamical properties of such mass are analyzed. We find that the first law of thermodynamics still works in such mass, and the global charge plays an important role in the relationship between the extra field and the work done by such extra field.

\section{Acknowledgments}
This research is supported by the National Natural Science Foundation of China under Grant Nos. 11273009 and 11303006.




\end{document}